\documentclass[a4paper]{PoS}
\usepackage[utf8]{inputenc}
\usepackage{amsmath}

\title{$a_1$ properties in hadronic tau decays}

\ShortTitle{$a_1$ properties in hadronic tau decays}

\author{\speaker{Ina Lorenz}\\%
       Department of Physics, Indiana University, Bloomington, IN 47405, USA\\
       Center for Exploration of Energy and Matter, Indiana University, Bloomington, IN 47408, USA\\
       E-mail: \email{ilorenz@indiana.edu}}

\author{Emilie Passemar\\
       Department of Physics, Indiana University, Bloomington, IN 47405, USA\\
       Center for Exploration of Energy and Matter, Indiana University, Bloomington, IN 47408, USA\\
       Theory Center, Thomas Jefferson National Accelerator Facility, Newport News, VA 23606, USA\\
       E-mail: \email{epassema@indiana.edu}}

\abstract{Hadronic tau decays belong to the processes that show a resonance-like structure in the axial vector current in the $1-2$ GeV range. This structure, often denoted as the $a_1$ meson, seems to show different properties in different processes. The process $\tau\rightarrow 3\pi\nu_{\tau}$ allows for a clean separation of weak and strong effects and a clear production mechanism. We examine how this structure can be related to interactions between the three pions that emerge in the final state. In particular we start from the interactions between all two body combinations.}

\FullConference{38th International Conference on High Energy Physics\\
          3-10 August 2016\\
         Chicago, USA}

\begin{document}

\section{Interest in $\tau\rightarrow3\pi\nu_{\tau}$}
\noindent The axial vector current in the few GeV range plays a role in many processes, for example decays of $B$ and $D$ mesons, Higgs decays and neutrino scattering. The models employed to describe this current partly boil down to summing up Breit Wigner lineshapes and fitting to the invariant mass of the final state system, e.g. in $\tau\rightarrow3\pi\nu_{\tau}$. The dominant resonance-like structure is denoted as $a_1$ meson. However, the properties of this meson, as listed in the PDG, depend on the production mechanism, and contrast significantly between $\pi P\rightarrow3\pi P$ and the tau decay. This situation calls for a conceptual improvement of the models. On one hand, of course the Dalitz plot distributions should be considered in any resonance model. On the other hand, more information on the spin structure is available from structure functions, observables that are directy related to helicity amplitudes. In order to use this additional information we refrain from any Breit Wigner parametrization and start from the two pion interactions which are known from $\pi\pi$ scattering.

\section{Definitions}
\begin{figure}\label{fig:feyn}
\center
\includegraphics[width=.6\textwidth]{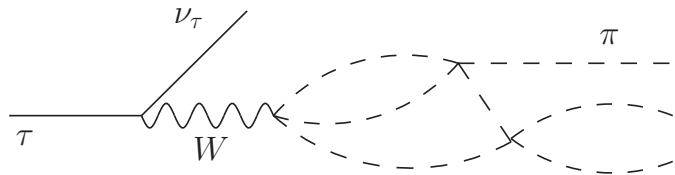}
\caption{Schematic rescattering of pions from the tau decay.}
\label{fig1}
\end{figure}

\noindent We consider the semileptonic decay, see Fig.~\ref{fig:feyn},
\begin{equation}\label{eq:taudec}
 \tau(l_1) \rightarrow \nu_{\tau}(l_2) + \pi(p_1) + \pi(p_2) + \pi(p_3),
\end{equation}
and its description following Refs.~\cite{Kuhn:1992nz, Colangelo:1996hs}. The general amplitude is
\begin{equation}
 \mathcal{M} = \cos\theta_C\frac{G_F}{\sqrt{2}}L_{\mu}H^{\mu}.
\end{equation}
where $\theta_C$ is the Cabibbo angle. The leptonic part can be written in the standard model as $L_{\mu} = \bar{u}(l_2)\gamma_{\mu}(1 - \gamma_5)u(l_1)$. For the hadronic part of the tau decay, we can write the generic matrix element
\begin{equation}\label{eq:matel}
 H_{\mu}^{ijkl} = \langle\pi^i(p_1)\pi^j(p_2)\pi^k(p_3)|V^l_{\mu}(0) - A^l_{\mu}(0)|0\rangle,\hspace{8pt} V^l_{\mu} = \frac{1}{2}\bar{q}\tau^l\gamma_{\mu}q,\hspace{8pt} A^l_{\mu} = \frac{1}{2}\bar{q}\tau^l\gamma_{\mu}\gamma_5q,
\end{equation}
where $ijkl$ are isospin indices and $\tau^l$ the Pauli matrices in isospin space. 
For an odd number of final state pions, the vector contribution vanishes and we consider only the axial part in the following. One possible decomposition is into form factors, as used e.g. in Ref.~\cite{Colangelo:1996hs}. However, we use instead a decomposition of the matrix element into helicity amplitudes, since these have a simple expansion into partial waves. The partial waves yield a simple form of the unitarity relations that we are interested in.\newline
The helicity amplitudes can be expressed in the Mandelstam plane where $s,t,u$ correspond to the two-by-two scattering element $\langle\pi(p_1)\pi(p_2)|\pi(-p_3)A_{\mu}(Q_{\mu})\rangle$. For three body decays often denoted as Dalitz plot invariants $s_1, s_2$ and $s_3$, here we use
\begin{equation}\label{eq:Q2}
 s = (p_1 + p_2)^2,\hspace{8pt} t = (p_2 + p_3)^2,\hspace{8pt} u = (p_1 + p_3)^2, \hspace{8pt} Q^2 = s + t + u - 3M_{\pi}^2.
\end{equation}
The center-of-mass scattering angle in each channel, $\theta_s, \theta_t$ and $\theta_u$, respectively, are related to the Kacser function  
\begin{equation}\label{eq:kacser}
 K(s) = \frac{t - u}{\cos\theta_s} = \sqrt{\lambda(s, M_{\pi}^2, M_{\pi}^2)}\sqrt{\lambda(s, Q^2, M_{\pi}^2)}.
\end{equation}
The Käll\'{e}n function $\lambda(a,b,c) = a^2 + b^2 + c^2 - 2(ab + bc + ca)$ can be written
\begin{equation}
 \lambda_{ab}(s) = \lambda(s, M_a^2, M_b^2) = [s - (M_a - M_b)^2][s - (M_a + M_b)^2].
\end{equation}
With the definitions $L_{\mu\nu} = L_{\mu}L_{\nu}^{\dagger}$ and $H_{\mu\nu} = H_{\mu}H_{\nu}^{\dagger}$ of the leptonic and hadronic tensor the differential decay rate is given by
\begin{equation}
 d\Gamma(\tau\rightarrow\nu_{\tau}3\pi) = \frac{1}{2m_{\tau}}|\mathcal{M}|^2d\Phi = \frac{G_F^2}{4m_{\tau}}\cos^2\theta_CL_{\mu\nu}H^{\mu\nu}d\Phi,\hspace{8pt}L_{\mu\nu}H^{\mu\nu} = \sum_X L_X W_X,
\end{equation}
where $d\Phi$ is the phase space element. $L_{\mu\nu}$ and $H_{\mu\nu}$, can be combined to form 16 symmetric and antisymmetric structure functions $W_X$.
One useful basis for the hadronic structure functions $W_X$ is defined via the polarization of the final state system. Consider the polarization vectors $\epsilon_{\mu}(\lambda)$ of the three pions in their c.m. frame or the W boson in its rest frame, respectively. We can now define the helicity amplitudes 
\begin{equation}
 \mathcal{A}^{ijkl}_{\lambda} := \langle\pi^i(p_1)\pi^j(p_2)\pi^k(p_3)|A^l_{\mu}(0)\epsilon^{\mu}(\lambda)|0\rangle, \label{eq:helamp}
\end{equation}
where the subscript denotes the helicity. The outgoing pions have the two possible physical states $|\pi_0\pi_0\pi_{\pm}\rangle$ and $|\pi_+\pi_-\pi_{\pm}\rangle$, that can be related by their isospin structure and crossing symmetry. In the following we will consider $\mathcal{A}_{\lambda}^{\pi_0\pi_0\pi_{\pm}}(s, t, u) \hat{=} \mathcal{A}_{\lambda}^{3311}(s, t, u)$ and neglect isospin breaking. 

\section{Method and parametrization}
\noindent We approximate the transverse amplitude similar to Refs.~\cite{Danilkin:2014cra, Guo:2015zqa},
\begin{align}\label{eq:exp}
 \mathcal{A}^{3311}_{+}(s, t, u) &\propto \sum_{l=0}^{l_{max}}\sum_I(2l+1)\bigg[d^l_{10}(\theta_s)\left(\frac{K(s)}{4s}\right)^{l-1} P_I^{3311}a_{+,Il}(s) + d^l_{10}(\theta_t)\left(\frac{K(t)}{4t}\right)^{l-1} P_I^{3131}a_{+,Il}(t) \notag\\
 &+ d^l_{10}(\theta_u)\left(\frac{K(u)}{4u}\right)^{l-1} P_I^{1331}a_{+,Il}(u)\bigg],
\end{align}
where $P_I^{ijmn}$ is the isospin projection operator. The relevant Wigner d-matrix is given by $d^l_{1 0}(\theta) = -\sin\theta/\sqrt{l(l+1)}P'_l(\cos\theta)$, where the prime denotes a derivative of the Legendre polynomial. The above expansion results in partial waves $a_{Il}$ that contain no kinematical but only dynamical cuts. This allows us to relate the parts of the partial waves that contain the left- and right-hand cuts $a_{Il}^{right/left}(s)$ in an iterative procedure suggested by Khuri and Treiman \cite{Khuri:1960zz}.\newline
In the following we always consider $a_{Il} = a_{+,Il}$. For each channel, we can write the discontinuity as a sum of the unitarity cut in this channel and those from the crossed channel as $a_{Il}^{left}$
\begin{align}
 \text{Disc }a_{Il}(s) = \rho(s)t_l^*(s)\left(a_{Il}^{right}(s) + a_{Il}^{left}(s)\right),
\end{align}
where $\rho(s) = \sqrt{1 - 4M_{\pi}^2/s}$ and $t_l(s)$ is the partial wave of the two-pion system, well-known from $\pi\pi$ scattering.
This discontinuity enters the standard dispersion relation, e.g. unsubtracted,
\begin{align}
a_{Il}^{right}(s) &= \frac{1}{\pi}\int_{s_0}^{\infty}ds'\frac{\text{Disc }a_{Il}^{right}(s')}{s'-s},\hspace{8pt} s_0 = 4M_{\pi}^2.
\end{align}
Expanding $\mathcal{A}^{3311}_{+}(s, t, u)$ in the $s$-channel physical region, comparing to Eq.~(\ref{eq:exp}), multiplying both sides with $P'_l(z_s)$ and integrating over $z_s = \cos\theta_s$ we can write
\begin{align}
 a_{Il}^{left}(s) &\propto \sum_{I',l'}(2l'+1)\int_{-1}^{+1}dz_s (1 - z_s^2) P'_l(z_s) \left(P'_{l'}(z_t)C^{II'}_{st}a_{I'l'}(t(s,z_s)) + P'_{l'}(z_u)C^{II'}_{su}a_{I'l'}(u(s,z_s))\right),
\end{align}
where $C_{st/su}$ are the standard crossing matrices, see e.g. Ref.~\cite{Guo:2015zqa}. To find a solution of this set of equations, we parametrize the transverse partial wave amplitudes similiar to Ref.~\cite{Colangelo:2016jmc}, as
\begin{align}\label{eq:drsol}
 a_{Il}(s) &= \Omega_{Il}(s)\left(\sum_i^{n-1} c_i s^i + \frac{s^n}{\pi}\int_{s_0}^{\infty}\frac{ds'}{s'^n}\frac{\rho(s')t^*_{l}(s')}{\Omega^*_{Il}(s')}\frac{a_{Il}^{left}(s')}{(s'-s)}\right),\hspace{2pt}\Omega_{Il}(s) = \exp\left(\frac{s}{\pi}\int_{s_0}^{\infty} \frac{ds'}{s'}\frac{\delta_{Il}(s')}{s'-s}\right),
\end{align}
where the Omn\`{e}s functions $\Omega_{Il}(s)$ contain the unitary cut in $s$, and we use their parametrization from Ref.~\cite{GarciaMartin:2011cn}. The term in brackets in Eq.~(\ref{eq:drsol}) contains the cuts from the crossed channels and corresponds to an $n$-times subtracted dispersion relation with the subtraction constants $c_i$. In a first step the left-hand cuts can be set to zero. However, three main restrictions of this approach are relevant in our case. First, the framework relies on the assumption that two body interactions dominate. This assumption is only justified at low energy, $Q^2 \ll $ 1 GeV$^2$. Second, the truncation of Eq.~$(\ref{eq:exp})$ induces an uncertainty that has to be tested in practice. Third, a precise knowledge of the individual waves decreases with increasing energy.

\section{Preliminary results}
\begin{figure}
\center
\includegraphics[width=.6\textwidth]{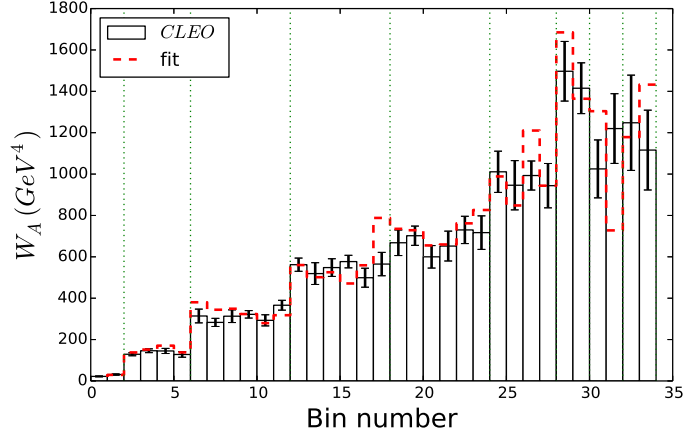}
\caption{Structure functions from CLEO \cite{Browder:1999fr}.}
\label{fig:WA}
\end{figure}

\begin{figure}
\center
\includegraphics[width=.6\textwidth]{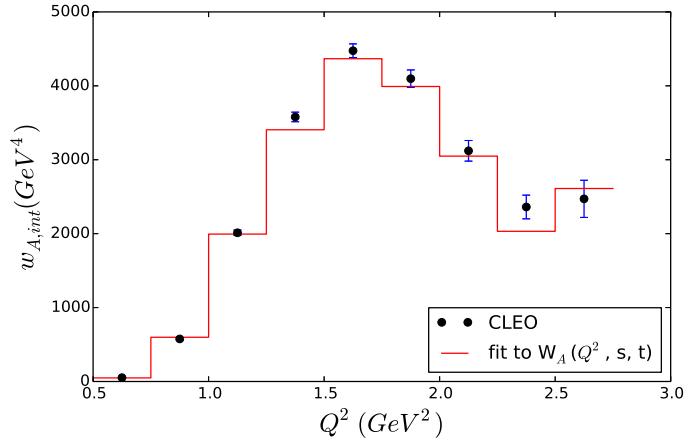}
\caption{Integrated structure functions from CLEO \cite{Browder:1999fr}.}
\label{fig:waint}
\end{figure}

\noindent Our calculation for the helicity amplitudes can directly be compared to the experimentally determined structure functions. All structure functions that are not compatible with zero according to CLEO \cite{Browder:1999fr} can be related to $W_A(s,t,u) \propto |\mathcal{A}^{3311}_+(s, t, u)|^2 + |\mathcal{A}^{3311}_-(s, t, u)|^2$ \cite{Kuhn:1992nz}. In Fig.~\ref{fig:WA} we show the structure function $W_A$ given by the CLEO collaboration in the corresponding bins and our fit result. Here we ignore the left hand cuts which corresponds to the first iteration step in a Khuri Treiman approach. For a complete analysis, see Ref.~\cite{Passemar16}. The dotted lines show the binning in $Q^2$, the solid line bars correspond to bins in $s$ and $t$ and the red dashed line to our preliminary fit.
Changing the variables by Eq.~(\ref{eq:Q2}) and integrating $W_A(Q^2, s, t)$ over $s$ and $t$ yields the integrated structure functions $w_{A, int}(Q^2)$ shown in Fig.~\ref{fig:waint}. Here, a three body resonance-like structure occurs and can be reproduced qualitatively by our parametrization based on two body interactions. Both figures on structures functions show a better agreement with the data for lower bin numbers or lower $Q^2$ values, respectively. For this kinematical region the Omn\`{e}s functions are known with higher precision. For close to vanishing $Q^2$ values the Khuri Treiman approach would be justified, as the dominating two body interaction corresponds to first order contributions in chiral perturbation theory.
The CLEO measurement \cite{Browder:1999fr} found the contributions from an off-shell $W$ to be compatible with zero. We thus approximate the decay rate by the transverse component \cite{Kuhn:1992nz}
\begin{align}
 \frac{dN}{NdQ^2}\bigg|_{\lambda=1} \propto \left(\frac{(M_{\tau}^2 - Q^2)}{Q^2}\right)^2\left(1 + 2Q^2/M_{\tau}^2\right)w_{A, int}(Q^2).
\end{align}
\begin{figure}
\center
\includegraphics[width=.6\textwidth]{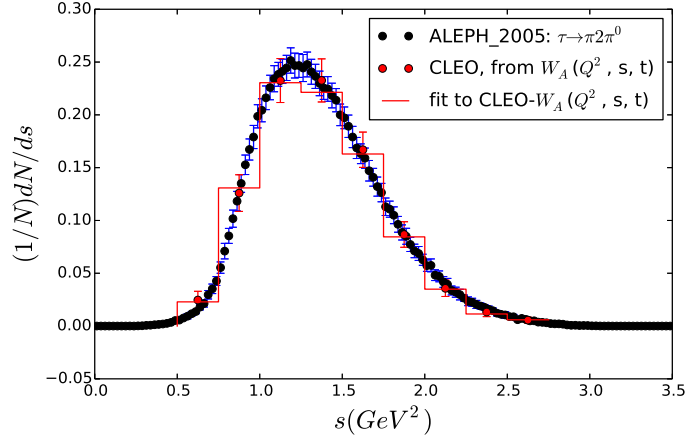}
\caption{Decay rate from a fit to CLEO structure functions, also compared to ALEPH \cite{Davier:2013sfa}.}
\label{fig:decay}
\end{figure}
\noindent The comparison to the decay rates from CLEO and ALEPH is given in Fig.~\ref{fig:decay}. Due to the very coarse grained bins in $s$ and $t$, we show the binning in $Q^2$. Again, the fit does not contain a specific parametrization of the three body resonance like a Breit Wigner, but merely two body interactions. This might hint towards an interesting origin of the $a_1$ meson, and/or towards the necessity for more iterations in the partial wave procedure or to include also three body unitarity. As a feasibility study, this work shows a good first description of the structure function and the tau decay rate. Therefore for a future detailed analysis it would be desirable to obtain the Dalitz plot distributions for a direct analysis, in particular from more precise measurements by Belle and BABAR. The full Dalitz plot information will help to separate the different uncertainties, namely the knowledge of the Omn\`{e}s functions, the range of applicability of the approach and the truncation error.

\bibliographystyle{JHEP}
\bibliography{tau}

\end{document}